\documentclass[sigconf, nonacm]{acmart}

\usepackage{xcolor}   
\usepackage{xspace}   
\usepackage{tabularray}
\usepackage{array}
\setcopyright{none}
\renewcommand\footnotetextcopyrightpermission[1]{} 

\def\hn{\sffamily\selectfont}
\newcommand{\mpfont}{\hn\scriptsize}

\ifx\noeditingmarks\undefined
  \newcommand{\MPworker}[2]{\unskip{\color{#1}\vrule\vrule}{\marginpar{\raggedright\color{#1}\mpfont #2}}}
  \newcommand{\pgwrapper}[3]{\begingroup \color{#1} #2: #3 \endgroup}
  
  \newcommand{\pgwrapperb}[1]{\textbf{#1}}
  \newcommand{\dangerwrapper}[1]{{\color{red}#1}}
  
\else
  \newcommand{\MPworker}[2]{\unskip}
  
  \newcommand{\pgwrapperb}[1]{}
  \newcommand{\pgwrapper}[3]{}
  \newcommand{\dangerwrapper}[1]{}
  
\fi

\newcommand{\zikai}[1]{\pgwrapper{green}{ZK}{#1}}
\newcommand{\ZK}[1]{\MPworker{green}{ZK: #1}}

\renewcommand{\paragraph}[1]{
\vspace{1ex}
\par
\noindent
\textbf{#1}} 

\newcommand{\heading}[1]{
\vspace{1ex}
\paragraph{#1}}

\newenvironment{myitemize2}%
{\begin{list}{\labelitemi}{\itemsep1pt \topsep2pt \parsep0.00in
          \partopsep=0pt \leftmargin1.2em}}%
          {\end{list}}

\let\latexusecounter=\usecounter
\def\compactsortof{\itemsep=0in \topsep=2pt \parsep=0.00in \partopsep=0pt \leftmargin=1.7em}
\newenvironment{myenumerate2}
{\def\usecounter{\compactsortof\latexusecounter}
  \begin{enumerate}}
    {\end{enumerate}\let\usecounter=\latexusecounter}

\newcommand{\robustness}{Robustness-$\delta$@K\xspace}
\newcommand{\robustnessatk}[2]{Robustness-{#1}@{#2}\xspace}
\newcommand{\robustnessat}[1]{Robustness-{#1}@10\xspace}

\newcommand{\recall}{Recall\xspace}
\newcommand{\recallat}[1]{Recall@{#1}\xspace}

\newcommand{\zilliz}{Zilliz\xspace}
\newcommand{\scann}{ScaNN\xspace}
\newcommand{\ivf}{IVFFlat\xspace}
\newcommand{\hnsw}{HNSW\xspace}
\newcommand{\puck}{Puck\xspace}
\newcommand{\diskann}{DiskANN\xspace}

\newcommand{\texttoimage}{Text-to-Image-10M\xspace}
\newcommand{\spacev}{MSSPACEV-10M\xspace}
\newcommand{\deep}{DEEP-10M\xspace}
\newcommand{\msmarco}{MSMARCO-10M\xspace}
\newcommand{\Deep}{DEEP-10M\xspace}
\newcommand{\Spacev}{MSSPACEV-10M\xspace}
\newcommand{\Texttoimage}{Text-to-Image-10M\xspace}
\newcommand{\marco}{MSMARCO\xspace}
\newcommand{\hotpotqa}{HotpotQA\xspace}

\newcommand{\wikipedia}{Wikipedia\xspace}

\newcommand{\qa}{Q\&A\xspace}

\newcommand{\Naiverag}{Naive RAG\xspace}
\newcommand{\agenticrag}{Agentic RAG\xspace}

\newcommand{\knn}{K-NN\xspace}
\newcommand{\recallatk}{Recall@K\xspace}

\begin{document}
\title{Towards \textit{Robustness}: A Critique of Current Vector Database Assessments} 

\author{Zikai Wang}
\affiliation{%
  \institution{Northeastern University}
}
\email{wang.zikai1@northeastern.edu}

\author{Qianxi Zhang}
\affiliation{%
  \institution{Microsoft Research}
}
\email{Qianxi.Zhang@microsoft.com}

\author{Baotong Lu}
\affiliation{%
  \institution{Microsoft Research}
}
\email{baotonglu@microsoft.com}

\author{Qi Chen}
\affiliation{%
  \institution{Microsoft Research}
}
\email{cheqi@microsoft.com}

\author{Cheng Tan}
\affiliation{%
  \institution{Northeastern University}
}
\email{c.tan@northeastern.edu}

%
%

\begin{abstract}
Vector databases are critical infrastructure in AI systems, and average recall is the
dominant metric for their evaluation. Both users and researchers rely on it to
choose and optimize their systems.

We show that relying on average recall is problematic. It hides
variability across queries, allowing systems with strong mean performance to
underperform significantly on hard queries. These tail cases confuse users and
can lead to failure in downstream applications such as RAG.

We argue that robustness---consistently achieving acceptable recall across
queries---is crucial to vector database evaluation.
We propose \emph{\robustness}, a new metric that captures the fraction of queries
with recall above a threshold $\delta$.
This metric offers a deeper view of recall distribution,
    helps vector index selection regarding application needs,
    and guides the optimization of tail performance.

We integrate \robustness into existing benchmarks and evaluate mainstream
vector indexes, revealing significant robustness differences.
More robust vector indexes yield better application performance, even with the same average recall.
We also identify design factors that influence robustness,
providing guidance for improving real-world performance.
\end{abstract}

\maketitle
\section{Introduction}
\label{s:intro}
\paragraph{A motivating example.}
Ana is a developer responsible for maintaining a Q\&A service.
The service relies on a vector database with
an average recall of 0.9---retrieving 90\% of the expected items on
average---when tested on the company's question-answering dataset.
One day, Ana learns about a new vector
database that runs faster. She configures the new database with the
dataset, tests it, and confirms that its average recall is also 0.9 but runs faster.
Satisfied, she deploys the new database to production.
However, the next day, users report difficulties in retrieving the
answers they expect. Ana is confused. She verifies that
the dataset, users' queries, and the recall metric remain unchanged.
Now, she wonders: what could have gone wrong?

This example illustrates a common issue in practice.
The root cause lies in the use of
\emph{average recall}: 
while average recall effectively communicates overall
performance, it fails to capture performance in the tail of the distribution. 
Tail performance, despite affecting only a small percentage of
cases, often disproportionately impacts end-user experiences in many applications~\cite{dean2013tail,arapakis2014impact,brutlag2009speed}.
Addressing this oversight is crucial and urgent, as vector databases
increasingly support critical functionalities in modern AI applications~\cite{pan2024vdbsurvey, wangMilvusPurposeBuiltVector2021, jing2024large}.

At the core of vector databases is nearest neighbor search, which aims
to find the most similar vectors (e.g., those closest in terms of Euclidean
distance) to the query from high-dimensional vector datasets.
%
%
Due to the \emph{curse of dimensionality}~\cite{clarkson1994algorithm}, performing
exact top-$K$ searches to find $K$ vectors closest to a given query is computationally
prohibitive for large-scale, high-dimensional datasets. Consequently, vector search relies on Approximate Nearest Neighbor Search (ANNS)~\cite{indyk1998approximate}, which aims
to maximize query recall 
by retrieving as many correct results as possible
within millisecond-level latency.
Numerous vector indexes~\cite{scann2024,jayaram2019diskann, hnsw, ivf,wangMilvusPurposeBuiltVector2021} have been developed to support efficient ANNS.


Existing evaluations of vector databases rely heavily on average recall,
defined as the proportion of ground-truth items among the top-$K$ returned results,
averaged across all queries.
While useful and intuitive, average recall masks query variability
and hides poor performance on hard cases.
As a result,
a vector index with high average recall can still be ``fragile'',
performing poorly on challenging queries and yielding inconsistent or unacceptable results for applications.

\begin{figure}
  \centering
  \includegraphics[width=\linewidth]{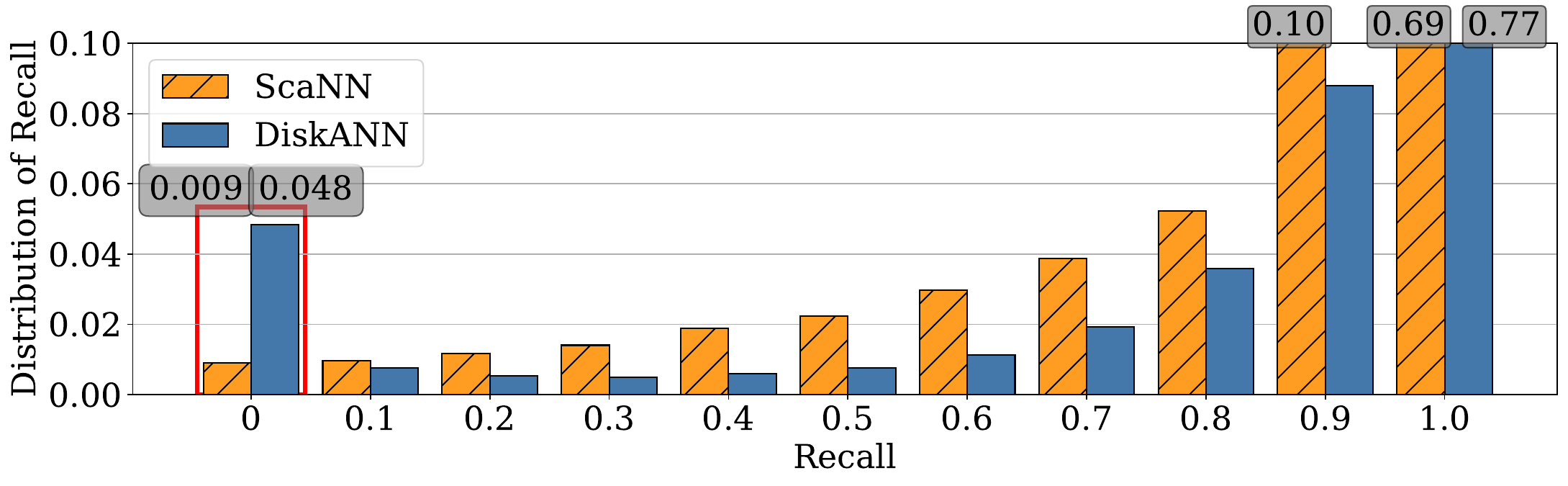}
  \caption{Recall distribution of \scann and \diskann on \marco, each achieving an average \recallat{10} of 0.9.
    Queries returning zero ground-truth items are highlighted with a red frame.
    \recallat{10}=0.9 and \recallat{10}=1.0 query results are shown above their bars.
    }
    \vspace{-4ex}
  \label{fig:recall-dist}
\end{figure}

Figure~\ref{fig:recall-dist} illustrates this issue:
we experiment on two popular vector indexes, \scann~\cite{scann2024} and \diskann~\cite{jayaram2019diskann},
using the widely-used Q\&A dataset \marco~\cite{msmarco}.
Both indexes achieve an average \recallat{10} of 0.9 on the query set.
However, the recall distributions vary widely,
with some queries achieving very low recall.
Notably, \diskann exhibits 4.8\% of queries with a recall of zero---an alarming number
that could translate to approximately 5\% of users receiving \emph{no} relevant answers
for their top-10 results.
\emph{This explains Ana's confusion:
even if average recall appears satisfactory,
the tail can lead to user frustration.}


Tail performance is crucial in many vector database applications.
In Retrieval-Augmented Generation (RAG)~\cite{lewis2020rag},
a large language model (LLM) can produce a correct answer if enough retrieved items are relevant,
but fails when few are~\cite{barnett2024seven}, the tail case exemplified by \diskann in Figure~\ref{fig:recall-dist}.
Similarly, users likely tolerate a small number of irrelevant search or recommendation items in their results,
but they will complain when only a small fraction of the results meet their expectations,
even if such cases are rare.

In addition, poor tail performance tends to compound in practice,
particularly in complex tasks requiring results aggregated across multiple
vector-indexed data sources~\cite{wang2024unims,chen2024onesparse} or multiple
rounds of interaction with the vector index~\cite{yang2024rag}.
For example, in multi-model RAG applications~\cite{wang2024unims},
a user query triggers multiple parallel ANN searches across indexes of different modalities,
where the final answer's quality is constrained by the poorest ANN search.
Similarly, 
in multi-hop RAG tasks such as Deep Research~\cite{openaideepresearch, geminideepresearch, opendeepsearch}, 
a single question involves multiple rounds of RAG, with each question in the
sequence conditioned on the answer from the preceding round. Failing to
retrieve relevant items in any round can propagate errors to subsequent
stages, thereby degrading overall performance~\cite{gao2024ragsurvey}.

The tail performance problem \emph{cannot} be solved by solely optimizing average recall.
As we will show later (\S\ref{s:eval}),
maximizing average recall 
does not always result
in a proportional improvement in low-recall queries.
Furthermore, efforts to improve average recall often uniformly increase the computational
burden across all queries, including those that are already performing well.
%
Finally, in high-dimensional spaces, data distributions are often skewed,
causing retrieval difficulty to vary across different regions of the query space.
Consequently, even with a high average recall, many indexes may
still exhibit low recall for certain queries when encountering hard-to-retrieve regions~\cite{aumuller2023recent,
aumuller2021role, li2020improving, wang2024boldsymbol, zhang2026adaef}.

The limitation of average recall highlights a deeper issue:
the community's ``obsession'' with this metric---reinforced by benchmarks like
\hspace{0pt}ANN-Benchmarks~\cite{aumuller2017ann, annbenchmark},
\hspace{0pt}Big-ANN-Benchmarks~\cite{simhadri2024results, bigann2023},
and recent large-scale evaluations~\cite{azizi2025graphann}---drives
efforts toward solely optimizing the average recall.
However, this focus potentially comes at the cost of tail performance
and may even hurt the performance of real-world applications.
As a result, this emphasis creates a disparity
between what researchers prioritize and the actual needs of real-world applications~\cite{pipitone2024legalbench}.

\bigskip



We argue that now is the time to establish a new metric that measures
tail performance and, arguably, redefines the core challenges for
the vector database community.
This metric must (a) distinguish tail performance across indexes;
(b) be application-oriented,
recognizing that different applications have distinct recall requirements in real-world scenarios;
and (c) be simple and intuitive---comparable to average recall---enabling users
and practitioners to clearly understand the bottlenecks in their vector indexes.
Note that simplicity is essential in practice:
comprehensive but complex metrics are difficult to adopt and deploy in practice.

In this paper, we introduce \emph{\robustness}, the first metric that satisfies all three criteria (a)--(c).
It quantifies the proportion of queries with recall
$\ge \delta$, 
where $\delta$ is an application-specific threshold.
For (a),
\robustness distinguishes tail performance:
in Figure~\ref{fig:recall-dist},
\scann achieves a \robustnessat{0.1} of 0.991,
while \diskann scores 0.952, a much lower value on \marco.
For (b),
it is parameterized by $\delta$, allowing alignment with application-specific recall requirements.
For (c),
\robustness is simple to interpret---as the
fraction of queries exceeding a recall threshold $\delta$---and
efficient to compute. 
We formally define \robustness in Section~\ref{s:def}.

Meanwhile, no prior metric meets all three criteria.
Mean Average Precision (MAP)~\cite{map} and Normalized Discounted Cumulative Gain (NDCG)~\cite{ndcg}
are average-based metrics that fail to capture tail behavior.
Mean Reciprocal Rank (MRR)~\cite{mrr} and percentile (e.g., 95th percentile \recall)
do not account for application-specific requirements.
We provide a quantitative comparison between \robustness and existing metrics
in Section~\ref{s:metric}.

To show the implication of \robustness in practice,
we evaluate it across six state-of-the-art vector indexes---\hnsw~\cite{hnsw},
\zilliz~\cite{wangMilvusPurposeBuiltVector2021},
\diskann~\cite{jayaram2019diskann}, 
\ivf~\cite{ivf}, \scann~\cite{scann2024}, \puck~\cite{puck}---using four representative datasets:
\Texttoimage~\cite{text2image}, \Spacev~\cite{msspacev}, \Deep~\cite{deep}, and \marco~\cite{msmarco}. 
Beyond evaluating indexes' performance on \robustness (\S\ref{ss:eval:overall}),
our experiments reveal several key findings that deepen the understanding of \robustness:
\begin{myitemize2}
    \item \textit{Trade-offs in index selection} (\S\ref{ss:eval:tradeoff}):
        \robustness highlights a new three-way trade-off when selecting vector indexes for targeted applications.
        Developers should holistically evaluate throughput, average
        recall, and \robustness to achieve optimal overall performance.

    \item \textit{Impact on end-to-end accuracy} (\S\ref{ss:eval:app}): 
        Differences in \robustness 
        of vector indexes significantly affect the end-to-end accuracy of applications such as RAG Q\&A.

    \item \textit{\robustness characteristics across indexes} (\S\ref{ss:eval:index}):
        The robustness of vector indexes varies significantly,
        even when their average recalls are the same.
        Notably, partition-based
        indexes (e.g., \ivf) tend to exhibit a more balanced recall
        distribution around the average recall, while graph-based indexes
        (e.g., \hnsw) show a more skewed distribution.
\end{myitemize2}
We summarize lessons learned in Section~\ref{s:lesson} with
our key observations,
guidelines for selecting vector indexes based on \robustness,
and several approaches to improve \robustness.

Our key contribution is proposing the new metric, \robustness, and
establishing its importance in vector search evaluation.
We strongly believe that \robustness will help improve vector databases
for applications in the new AI era.


\section{Background and Motivation}
\label{sec:background}

\subsection{Vector Search}

Vector search is becoming increasingly important in the modern AI paradigm.
In particular, deep learning encodes data from various domains---text, images, and speech---into
high-dimensional vector representations, typically ranging from tens to
thousands of dimensions~\cite{andoni2018approximate}, enabling advanced semantic understanding and
analysis~\cite{mitra2018semantic}. Vector search on these datasets facilitates a wide range of
AI applications, including semantic search~\cite{mitra2018semantic}, recommendation~\cite{cen2020controllable},
and retrieval-augmented generation (RAG)~\cite{jing2024large}.

In essence, vector search identifies the
$K$ nearest neighbors (KNN) of a given query within the vector dataset, where
both the vector dataset and the query are embeddings produced by deep learning models.
Given a dataset $\mathcal{X} \in \mathbb{R}^{n\times d}$ consisting of $n$
vectors in a $d$-dimensional space, KNN identifies the $K$ closest vectors
to a query vector $x_q \in \mathbb{R}^d$ based on a distance metric, such as
Euclidean or cosine distance, where $K$ is a predefined parameter.

\paragraph{Approximate Nearest Neighbor Search (ANNS).}
Due to the curse of dimensionality~\cite{bellman1966curse, indyk1998approximate}, computing exact results on
large-scale vector dataset requires substantial computational cost and high
query latency. As a result, vector search often relies on Approximate Nearest
Neighbor Search (ANNS), which sacrifices some accuracy to achieve approximate
results with significantly reduced computational effort, typically completing
in milliseconds, thus enabling support for online applications.

\paragraph{Recall@K.}
Search accuracy is typically evaluated using \emph{recall@K}, which measures
the proportion of relevant results retrieved by an ANNS query. Specifically,
\recallat{K} is defined as 
\[
Recall@K = \frac{|C \cap G|}{K},
\]
where $C$ represents
the set of results returned by the ANNS and $G$ is the set of ground truth
results returned by KNN. Both sets contain exactly $K$ elements.
Recall effectively captures how closely the results returned by an ANNS align
with the ground truth. 

\subsection{Vector Index}
\label{background:index}
\label{ss:index}

\begin{figure}[t]
    \centering
    \includegraphics[width=0.8\columnwidth]{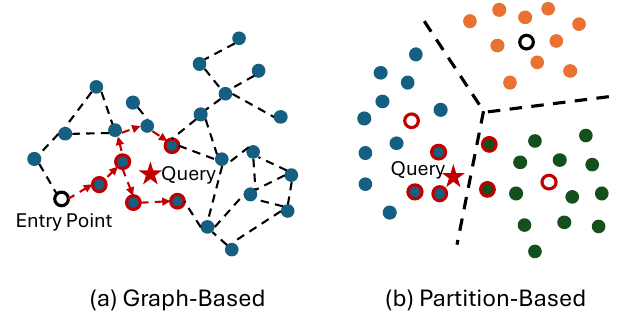}
    \caption{
 Overview of a graph-based index (left) and a partition-based index (right).
    In both cases, the query is represented as a red star, and dataset points are
    shown as blue, orange, and green dots, with dots bordered in red indicating the
    top 5 nearest neighbors to the query.
    In the graph-based index (a),
    dashed lines represent edges between vectors in the graph. A hollow dot
    indicates the entry point of the graph search, while red arrows trace the search path.
    In the partition-based index (b), each
    color corresponds to a distinct partition, and dashed lines denote partition
    boundaries. Hollow dots represent partition centroids, with those bordered in
    red being the top 2 nearest centroids to the query.
    %
    }
    \label{fig:index}
\end{figure}

Vector indexes support efficient ANNS by organizing the data in a
way that allows the search process to access only a small subset of the data to
obtain approximate results.
Currently, there are two major categories of vector indexes,
as shown in Figure~\ref{fig:index}.


\paragraph{Graph-based index.}
A graph $G = (V, E)$ is used to organize vector data, where each vertex
$v \in V$ represents a vector, and an edge $e \in E$ is created between two vertices
if their corresponding vectors are sufficiently close in the vector space~\cite{hnsw}.
For a graph-based index,
the number and arrangement of edges are key design considerations,
as they significantly influence the efficiency and navigability of the graph index.
In particular,
many approaches~\cite{hnsw,jayaram2019diskann} introduce a parameter \emph{efConstruction} (or \emph{M} in \diskann~\cite{jayaram2019diskann}) 
to determine how many edges each vertex is connected to.
Typically, \emph{efConstruction} is set to tens or hundreds,
with larger \emph{efConstruction} values improving recall at the
cost of higher storage and computational overhead.

For a graph-based index,
the search process usually begins at one or more predefined entry points and
proceeds by greedily traversing to the neighbor of the currently visited vertex
that is closest to the query. This traversal continues until the algorithm
determines that the nearest results have been identified.
The method proposed in \hnsw~\cite{hnsw} and \diskann~\cite{jayaram2019diskann} employs a priority queue
of size \emph{efSearch} (or \emph{Ls} in \diskann~\cite{jayaram2019diskann}) 
to store the current nearest results. During traversal, newly
accessed vectors are assessed and, if appropriate, inserted into the priority
queue. The traversal process concludes when the priority queue no longer
updates, and the $K$ nearest results are subsequently retrieved from the queue.

\paragraph{Partition-based index.}
Partition-based indexes~\cite{ivf,scann2024, puck} divide data into $n\_list$ 
partitions based on locality, commonly employing methods like $k$-means
clustering. Each partition is represented by a representative vector, such as a
centroid. During a search, the process first identifies some $n\_probe$ closest
representative vectors, where both $n\_list$ and $n\_probe$ are configurable parameters, and
then retrieves data from their corresponding partitions to determine the $K$
nearest results.
ScaNN~\cite{scann2024} and Puck~\cite{puck} also employ quantization methods
to compress the vectors within each partition. At the final stage of the
search, the original, precise vectors are used to re-rank the results obtained
from the partitions.


\subsection{Benchmarks and Metrics in Practice}

\paragraph{Benchmark.}
Two benchmarks are commonly used to evaluate vector indexes.
\emph{ANN-Benchmarks}~\cite{annbenchmark} provides a standardized methodology
for evaluating various indexes and parameter configurations.
\emph{Big-ANN-Benchmarks}~\cite{bigann2023} extends this approach to
larger-scale datasets up to 1 billion vectors, and supports a broader range of scenarios.
Both benchmarks build indexes on a given vector dataset and evaluate them using a corresponding query set.

\paragraph{Metric.}
The benchmarks measure accuracy using \emph{average recall} of all queries in
the query set and throughput using \emph{Queries Per Second (QPS)}.
For a given index, achieving higher average recall typically requires more computation,
resulting in lower QPS, whereas lower average recall leads to higher QPS,
illustrating a trade-off. This relationship is often visualized through a
performance curve, with comparisons focusing on the QPS achieved at a given
level of average recall.
Applications use these benchmarks to evaluate candidate indexes and select
suitable parameters, while the research community leverages them to
guide new indexes.
Other metrics have been proposed in the academic literature,
which we discuss in detail in Section~\ref{s:metric}.

\subsection{Motivation: Average Recall Falls Short}
\label{background:limitation}



Average recall is commonly used to measure the accuracy of vector indexes.
However, it falls short of capturing practical performance, as it focuses
solely on the average and largely ignores tail performance. Tail performance is
crucial in real-world applications, as demonstrated in Ana's case (\S\ref{s:intro}).

Average recall is insufficient to capture the tail
because high-dimensional data distributions are typically skewed rather than uniform.
Prior work~\cite{wang2024boldsymbol} has shown that
query difficulty varies across different regions of the vector space. This
variation means that queries in distinct areas can exhibit significantly
different recall values, even within the same index. Consequently, average
recall provides an incomplete view of the recall distribution. Indexes
achieving high average recall may still perform poorly for certain queries,
making average recall an inadequate metric for evaluating practical
performance.

\paragraph{The impact on real-world applications.}
Applications often assume that vector indexes achieve a baseline level of recall
to function as expected.
This is supported by prior research~\cite{zhao2024towards, pipitone2024legalbench},
showing a strong correlation between retrieval recall and application correctness.
For example, Zhao et al.~\cite{zhao2024towards} examined RAG applications across diverse datasets,
finding that retrieval recall requirements vary widely, from 0.2 to 1.0, depending on
the specific needs of the application.
Queries with low recall (e.g., $<$0.2) often fail to retrieve critical information,
directly undermining the correctness of application outputs.

However, this application's requirements of recall$\ge$0.2 cannot be captured by today's metrics.
As illustrated in Figure~\ref{fig:recall-dist} (\S\ref{s:intro}),
even when \diskann achieves a high average \recallat{10} of 0.9,
5.72\% of queries still have \recallat{10} $<$0.2.
These low-recall queries make it nearly impossible to support accurate
application results, highlighting the limitations of relying solely on average
recall as an evaluation metric.
This challenge is further exacerbated in applications requiring multiple
retrieval operations for a single request,
such as Deep Research~\cite{openaideepresearch, geminideepresearch, opendeepsearch}
and others~\cite{chen2024onesparse,wang2024unims,yang2024rag,jeong2024adaptive,press2022measuring,trivedi2022interleaving}.


\paragraph{End-to-end applications: RAG.}
Identical average recall does not guarantee comparable application performance.
To show this,
we evaluate two RAG applications, a naive Q\&A task on \marco and an agentic multi-hop task on \hotpotqa (\S\ref{evaluation:application}).
In both cases, partition-based indexes consistently outperform graph-based indexes at the same average recall.
For example, in the naive Q\&A task,
a \scann configuration with \recallat{10}=0.90 achieves the same end-to-end accuracy
as a \diskann configuration with \recallat{10}=0.96.
In the agentic task,
\ivf at average \recallat{5}=0.85 matches the accuracy of \hnsw at \recallat{5}=0.9.
These results highlight that
the same average recall can yield very different application performance,
requiring a more precise evaluation metric.


In conclusion, a new metric is needed to better characterize the recall
distribution---specifically to address application-specific recall requirements and
evaluate the robustness of vector indexes.

\section{Robustness Definition}
\label{s:def}

In this section, we formally define \robustness.

\paragraph{Dataset and query set.}
%
Our robustness definition aligns with the setup for average recall---it is
measured using a dataset and a corresponding query set. A vector index is
constructed from the dataset and evaluated using the query set.

\noindent
\textit{Dataset:} A \textit{dataset} $\mathcal{X}$ is a collection of $n$ data
points, where each data point $x_i$ is represented as a vector in a
$d$-dimensional space: 
\[
\mathcal{X} = \{x_1, x_2, \dots, x_n\} \quad \text{where} \quad x_i \in \mathbb{R}^d.
\]

\noindent
\textit{Query Set:} A \textit{query set} $\mathcal{Q}$ is a collection of $m$
query points, where each query point $q_j$ is represented as a vector in the
same $d$-dimensional space as the dataset:
\[
    \mathcal{Q} = \{q_1,q_2,\dots, q_n\} \quad \text{where} \quad q_i \in \mathbb{R}^d.
\]

\paragraph{Vector index and ANN search.}
A vector index $\mathcal{I}$ is constructed on a given dataset using parameters $\Theta$.
Different vector indexes have different construction processes, each with
different parameters to tune (\S\ref{ss:index}):
\[
    \mathcal{I} \gets \textsc{IndexConstruction}(\mathcal{X}, \Theta).
\]

\noindent
An approximated nearest neighbor (ANN) search
aims to retrieve $K$ vectors from the dataset $\mathcal{X}$
that are closest to a given query vector $q \in \mathcal{Q}$,
based on the runtime parameters $\theta$ of the index $\mathcal{I}$:
\[
r = \mathcal{I}(\theta, q, K).
\]
Here, $r \in \mathbb{R}^{K \times d}$ represents an array of size $K$, where each element is a vector in $R^d$.


\paragraph{Recall.}
For an ANN query $r = \mathcal{I}(\theta, q, K)$,
its recall is defined as:
\[
    R = \frac{|r \cap \textsc{KNN}(q,\mathcal{X},K)|}{K},
\]
where $\textsc{KNN}(\cdot)$ denotes the $K$-nearest neighbor function,
which provides the ground truth set of the $K$ closest vectors to $q$ in $\mathcal{X}$.
The operator $|\cdot|$ returns the number of items in the set,
and $r \cap \textsc{KNN}(q,K,\mathcal{X})$ represents the intersection of the retrieved result $r$
and the ground truth set.

\paragraph{\robustness.}
We define \robustness as follows:
\[
\text{Robustness-}\delta@K = \frac{1}{m} \sum_{i=1}^{m} \mathbb{I}\left( R_i \geq \delta \right),
\]
where $m$ is the size of query set $\mathcal{Q}$;
$R_i$ is the recall of query $q_i$;
and \( \delta \) represents the required recall threshold for each query.
The function \(\mathbb{I}(\cdot)\) is the indicator function:
\[
\mathbb{I}\left( R_i \geq \delta \right) = 
\begin{cases}
1, & \text{if } R_i \geq \delta, \\
0, & \text{otherwise}.
\end{cases}
\]

\paragraph{Implication of \robustness.}
\robustness is designed to be application-oriented,
with \(\delta\) representing the minimum or expected recall
for any query \(q_i \in \mathcal{Q}\) that applications assume or accept.
For the previously discussed RAG application~\cite{zhao2024towards},
a recall$\ge$0.2 is a minimum threshold needed to produce high-quality answers.
In this case, users should evaluate the system using $\delta=0.2$.

\section{Comparison to existing metrics}
\label{s:metric}
\label{s:metrics}

Beyond average recall, several other metrics have been proposed in the
literature~\cite{map,ndcg,mrr}.
We compare \robustness with these metrics byand show that it
captures distinct index characteristics not reflected by the others.
\robustness does not subsume the other metrics;
rather, they offer complementary perspectives.

\heading{Common metrics in information retrieval (IR).}
Several standard metrics are used to evaluate retrieval effectiveness:
\begin{myitemize2}

    \item \emph{MAP@K}~\cite{map}: Mean Average Precision at K computes the
        average precision over the top-K retrieved results.
        For each query, it calculates the precision at the rank of
        each relevant retrieved item (if present in the top-K ground truth),
        and averages these values.
        MAP@K is the mean of these per-query average precisions across all queries.

    \item \emph{NDCG@K}~\cite{ndcg}: Normalized Discounted Cumulative Gain at K
        accounts for the rank positions of relevant items in the retrieved list.
        For each query, DCG@K is computed as the sum of
        $1/\log(\text{rank} + 1)$,  where rank
        is the position of a retrieved item in the ground-truth list.
        NDCG@K is obtained by normalizing DCG@K by IDCG@K, the ideal DCG when all top-K
        results are perfectly ranked.
        The final score is the mean NDCG@K across all queries.

    \item \emph{MRR@K}~\cite{mrr}: Mean Reciprocal Rank at K evaluates the rank
        of the first relevant result.
        For each query, it computes the reciprocal of the rank of the first retrieved
        item that appears in the top-K ground-truth set.
        MRR@K is the mean of these reciprocal ranks over all queries.
\end{myitemize2}

\heading{Systematic comparison across datasets.}
To assess whether these metrics add information beyond average recall,
we compute the Pearson $r^2$ between each metric and average \recallat{10}
across all index configurations within the evaluation recall range (0.70--0.95)
on each of the four benchmark datasets
(\texttoimage, \spacev, \deep, and \msmarco).
A high $r^2$ ($\approx 1.0$) means the metric is a near-linear function
of average recall and offers no additional information.
A lower $r^2$ indicates the metric captures aspects of performance that
average recall does not.

\begin{figure}
\centering
\includegraphics[width=0.95\linewidth]{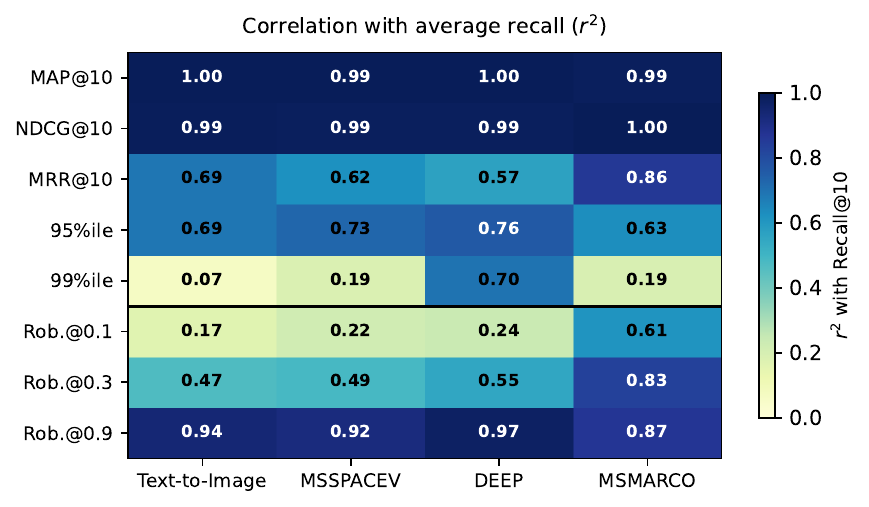}
\vspace{-2ex}
\caption{Correlation ($r^2$) between each metric and average \recallat{10},
computed across all index configurations on four datasets.
Green (low $r^2$): the metric captures information beyond recall.
Red (high $r^2$): the metric is redundant with recall.
}
\label{fig:metric_r2}
\vspace{-2ex}
\end{figure}

Figure~\ref{fig:metric_r2} presents the results.
MAP@10 and NDCG@10 have $r^2 > 0.98$ on all four datasets,
confirming that they are near-linear functions of average recall and add no
information for distinguishing indexes.
This is expected: both metrics aggregate per-query scores by averaging,
which, like average recall, masks the distribution across queries.

MRR@10 shows moderate correlation ($r^2 = 0.54$--$0.86$).
MRR is sensitive to zero-recall queries (which contribute MRR$=$0),
so it partially reflects the same tail failures that \robustness captures.
However, MRR also varies with the rank of the first correct result,
introducing noise unrelated to the recall distribution.
As a result, MRR is not consistently informative across datasets.

In contrast, \robustness at low $\delta$ values
captures substantial information beyond recall.
Across all four datasets, \robustnessat{0.1} achieves $r^2 = 0.17$--$0.61$,
meaning a large fraction of its variance is \emph{not} explained by average recall. This is crucial for identifying indexes that perform poorly on challenging queries.
\robustnessat{0.3} similarly shows $r^2 = 0.45$--$0.83$,
well below the near-perfect correlation of MAP and NDCG.
At higher $\delta$ values (e.g., $\delta=0.9$), \robustness correlates more
strongly with recall ($r^2 > 0.87$), which is expected:
high-$\delta$ robustness measures how many queries achieve near-perfect recall,
naturally tied to the overall recall level.
Percentile metrics show inconsistent $r^2$ across datasets
(Figure~\ref{fig:metric_r2}); we analyze the reasons below alongside
their conceptual limitations.

\heading{What does the residual capture?}
To understand what \robustness reveals that other metrics do not,
we compare graph-based and partition-based indexes at the same average recall
($\approx 0.9$, $\pm 0.02$) on \texttoimage.
At this fixed recall, MAP@10, NDCG@10, and MRR@10 differ by less than 1\%
between the two index families, seeing both as essentially identical.
However, the \emph{failure rate} (the fraction of queries with recall below
$\delta$, i.e., $1 - \text{Robustness-}\delta$) reveals dramatic differences:
graph-based indexes have a $5.4\times$ higher failure rate at $\delta=0.1$
and $3.3\times$ higher at $\delta=0.3$ compared to partition-based indexes.
In absolute terms, graph-based indexes fail to return \emph{any} relevant result
for 2.1\% of queries, versus 0.4\% for partition-based indexes, a gap entirely
invisible to MAP, NDCG, or MRR.
The pattern is consistent across all four datasets
($3.4\times$--$7.6\times$ at $\delta{=}0.1$).

\heading{Compared to percentile.}
A common approach to evaluating tail performance is to use percentiles.
Many systems report 95\%ile latency to characterize tail behavior,
and the same idea can be applied to recall.
However, percentile-based recall metrics suffer from two limitations.

First, percentiles are \emph{statistically unreliable} for discrete metrics.
With $K{=}10$, per-query recall takes only 11 discrete values,
and the percentile must be one of these.
As shown in Figure~\ref{fig:metric_r2}, the 99\%ile $r^2$ ranges
from $0.05$ on \texttoimage to $0.67$ on \deep.
On harder datasets, the 99\%ile collapses to 0.0 for over 60\% of
configurations regardless of average recall (low $r^2$ from discretization noise);
on easier datasets, it tracks recall smoothly but becomes redundant (high $r^2$).
\robustness avoids this trade-off through the tunable threshold $\delta$:
lower $\delta$ values probe the tail on hard datasets, while higher values
remain informative on easier ones.

Second, percentiles are \emph{distribution-oriented} rather than application-oriented:
they describe what the system delivers, not what the user requires.
In practice, choosing which percentile to report is often arbitrary
and difficult to justify.
Consider the Q\&A RAG application built on either \scann or
\diskann. Both indexes achieve the same average recall. Based on the commonly
used 95\%ile recall (sorting per-query recall in descending order and
selecting the 95th percentile), \diskann appears better than \scann (0.7 vs.
0.6). However, in practice, \diskann yields lower end-to-end accuracy for the
application (0.740 vs. 0.758). This example illustrates that
the index with (arbitrarily chosen) higher percentile recall (e.g., 95\%ile)
may underperform in application accuracy.

\emph{We argue that the new metric should be application-oriented.}
Different AI applications have different recall requirements~\cite{zhao2024towards, pipitone2024legalbench,wang2023uncertaintyquantificationfairnesstwostage}.
As we will show later (\S\ref{evaluation:application}),
the Q\&A example requires recall@10 $\ge$ 0.2 for useful final results.
Thus, regardless of how strong the 95\%ile recall is,
what matters is ensuring fewer queries fall below the 0.2 threshold.
\robustness is explicitly designed to be application-oriented:
$\delta$ represents the critical recall threshold required by the application.
For the Q\&A example above, \scann's \robustnessat{0.2} exceeds that of \diskann (0.997 vs. 0.974),
aligning with the application's final accuracy.
\section{Experimental Evaluation}
\label{sec:evaluation}
\label{s:eval}

We answer the following questions:
\begin{myitemize2}
    \item 
        How do state-of-the-art vector indexes perform under the new metric, \robustness? (\S\ref{ss:eval:overall})


    \item How can \robustness assist in index selection,
        and what trade-offs should users consider? (\S\ref{ss:eval:tradeoff})

    \item 
        What is the correlation between \robustness and the end-to-end performance of applications?
        (\S\ref{ss:eval:app})

    \item 
        What are the robustness characteristics of different families of indexes?
        (\S\ref{ss:eval:index})

\end{myitemize2}

\begin{figure}[b]
    \centering
    \resizebox{\columnwidth}{!}{%
    \begin{tabular}{@{}l|ccccc@{}}
        \toprule
         & Data type & Dimension & Dataset size & Query set & Distance \\
        \midrule
        \texttoimage & float32 & 128 & 10M & 100K & inner product  \\
        \spacev & uint8 & 100 & 10M & 29K & L2 \\
        \deep  & float32 & 96 & 10M & 10K & L2  \\
        \marco & float32 & 768 & 8.8M & 6.9K & inner product \\
        \bottomrule
    \end{tabular}
    }
    \caption{Dataset characteristics.
    }
    \label{tab:dataset}
    \vspace{-2ex}
\end{figure}

\begin{figure*}[t]
    \centering
    \label{tab:overall-parameters}
    \resizebox{\textwidth}{!}{%
    \footnotesize
    \begin{tblr}{
        row{1} = {c},
        row{2} = {c},
        cell{1}{1} = {r=2}{},
        cell{1}{2} = {r=2}{},
        cell{1}{3} = {c=4}{},
        cell{3}{1} = {c},
        cell{3}{2} = {r=3}{c},
        cell{4}{1} = {c},
        cell{5}{1} = {c},
        cell{6}{1} = {c},
        cell{6}{2} = {r=3}{c},
        cell{7}{1} = {c},
        cell{8}{1} = {c},
        hline{1,3,6,9} = {-}{},
        hline{2} = {3-6}{},
        hline{4-5,7-8} = {1,3-6}{},
      }
      Index   & Type  & Parameters                                                  &                                                   &     &                                                                                                                                                    \\
              &       & \texttoimage                                                & \spacev                                           & \deep & MSMARCO \\
      {\\\hnsw\\}   & Graph & {M=32, efConstruction=300,\\efSearch=64 to 512}           & {M=32, efConstruction=300,\\efSearch=20 to 300} & {M=16, efConstruction=300,\\efSearch=20 to 300}   & {M=16, efConstruction=500,\\efSearch=30 to 500}                                                                                          \\
      {\diskann} &       & {R=64, L=500, Ls=20 to 400}                             & {R=32, L=300, Ls=10 to 250}                   & {R=16, L=300, Ls=10 to 250}  & {R=32, L=500, Ls=15 to 85}                                                                                                                \\
      {\zilliz}  &       & {R=48, L=500, Ls=20 to 200}                             & {R=32, L=500, Ls=10 to 100}                   & {R=16, L=300, Ls=10 to 80}   & {--}                                                                                                               \\
      {\ivf} & Partition   & {n\_list=10000, n\_probe=10 to 80}                         & {n\_list=10000, n\_probe=10 to 200}             & {n\_list=10000, n\_probe=6 to 40}  & {n\_list=4000, n\_probe=10 to 200}                                                                                                  \\
      {\\\scann\\}   &       &   {\#leaves=40000, \\ro\_\#n=30 and 150, \\\#l\_search=10 to 100}      &        {\#leaves=40000, \\ro\_\#n=50 and 150, \\\#l\_search=10 to 200}  &   {\#leaves=40000, \\ro\_\#n=50 and 150, \\\#l\_search=10 to 200}  &   {\#leaves=10000, \\ro\_\#n=150, \\\#l\_search=10 to 120} \\
      {\puck}    &       & {s\_\#c=10, s\_range=30 to 350}   &    {s\_\#c=10, s\_range=10 to 100} &   {s\_\#c=10, s\_range=15 to 90}   &   {s\_\#c=10, s\_range=30 to 350}                                                         \\
    \end{tblr}}%
    \caption{Index parameters for all experiments.
    \emph{For graph-based indexes},
    M and R represent the maximum degree of a node;
    efConstruction and L represent the search list length during building;
    efSearch and Ls represent the search list length during searching.\\
    \emph{For partition-based indexes},
    n\_list and \#leaves represent the number of clusters;
    n\_probe and \#l\_search represent the number of clusters searched.
    In \scann, ro\_\#n is short for reorder\_num\_neighbors. It represents the
    number of KNNs to be reranked. \#leaves is short for num\_leaves, and
    \#l\_search is short for num\_leaves\_to\_search.
    In \puck, s\_\#c represents the number of coarse, and s\_range is short for tinker\_search\_range, which represents the number of finer clusters searched. Zilliz is excluded from MSMARCO because the Docker image has a bug quantizing 768d vectors (unfixable without vendor patch).
    }
    \label{tab:parameter}
\end{figure*}

\paragraph{Vector indexes.}
We experiment with six state-of-the-art indexes:
\begin{myenumerate2}

    \item \emph{\hnsw}~\cite{hnsw}: A popular graph-based index
        supported by many modern vector
        databases~\cite{pgvector,wangMilvusPurposeBuiltVector2021,
        zhang2023vbase, alibaba, aws, douze2024faiss}.

    \item \emph{\diskann}~\cite{jayaram2019diskann}: A disk-based graph
        index achieving state-of-the-art performance on billion-scale vector
        datasets and integrated into multiple vector
        databases~\cite{azure,wangMilvusPurposeBuiltVector2021}.
        We use the in-memory version of \diskann.

    \item \emph{\zilliz}~\cite{hnsw}: A commercial graph-based index built on
        \diskann and \hnsw. \zilliz ranked first or second across all tracks in
        the Big-ANN-Benchmarks 2023~\cite{bigann2023}. 
        We use hnsw as the underlying index for \zilliz.

    \item \emph{\ivf}~\cite{ivf}: A widely used partition-based index
        supported by many modern vector
        databases~\cite{pgvector,wangMilvusPurposeBuiltVector2021,
        zhang2023vbase, alibaba, aws, douze2024faiss}.

    \item \emph{\scann}\cite{scann2024}: A highly optimized partition-based
        index with quantization. \scann achieves state-of-the-art performance
        in two tracks of the Big-ANN-Benchmarks 2023 to which it is
        applicable~\cite{sun2024soar}, as well as in the ANN-Benchmarks on the
        GloVe-100-angular dataset.

    \item \emph{\puck}~\cite{puck}: A hybrid index that combines
        multi-level partitioning with a graph-based refinement step (tinker).
        \puck demonstrated the best performance across multiple datasets in the
        Big-ANN-Benchmarks 2021 competition track.

\end{myenumerate2}

\paragraph{Datasets.}
We perform experiments on four well-known vector datasets:
\begin{myitemize2}

    \item \emph{Text-to-Image}~\cite{text2image}: A dataset derived from Yandex
        visual search, comprising image embeddings generated by the
        Se-ResNext-101 model~\cite{hu2018squeeze} and textual query embeddings
        created using a variant of the DSSM model~\cite{huang2013learning}. The
        two modalities are mapped to a shared representation space by
        optimizing a variant of the triplet loss, leveraging click-through data
        for supervision.

    \item \emph{MSSPACEV}~\cite{msspacev}: A production dataset derived from
        commercial search engines, comprising query and document embeddings
        generated using deep natural language encoding techniques. These
        embeddings capture semantic representations, enabling effective
        similarity search and retrieval.

    \item \emph{DEEP}~\cite{deep}: An image vector dataset generated using the
        GoogLeNet model, pre-trained on the ImageNet classification task. The
        resulting embeddings are subsequently compressed via PCA to reduce
        dimensionality while preserving essential features for effective
        similarity search.

    \item \emph{MSMARCO}~\cite{msmarco}: A large-scale passage retrieval dataset
        from Microsoft Bing. We encode the corpus and queries using the
        LLM-Embedder~\cite{zhang2023llmembedder} model, producing 768-dimensional
        embeddings with inner product similarity.

\end{myitemize2}
Figure~\ref{tab:dataset} provides detailed characteristics of these datasets.
The index configurations we used are the in-memory versions of the indexes, so we limit the dataset size to fit the memory.
We use the 10M version of the first three datasets provided by the Big-ANN-Benchmarks~\cite{bigann2023},
which we call \texttoimage, \spacev, and \deep.
\marco contains 8.8M passages with 7.0K queries. 

\paragraph{Setup.}
We run experiments on a machine with dual Intel(R) Xeon(R) Gold 6248R CPUs, each with 24 cores at 3.00GHz, 
with 1.5TB memory and Ubuntu 24.04.
of memory and Ubuntu 24.04.
For all experiments,
we use Docker version 28.1.1 with Python 3.13.5. 
%
We run all the experiments with 16 threads
We extend the Big-ANN-Benchmark framework by integrating the \robustness as an evaluation metric.
Alongside \robustness, the Big-ANN-Benchmark framework originally provides Query Per Second (QPS) and average recall.
In the rest of our evaluation, we will use average recall@10 to represent the average recall for the top-10 ANN search.



\subsection{Index Performance for \robustness}
\label{evaluation:overall}
\label{ss:eval:overall}

We comprehensively evaluate existing vector indexes following standard benchmark configurations to show their robustness values.
The operating points (i.e., index configurations) 
are selected based on the
documentation of indexes and their configurations in the
Big-ANN-Benchmark~\cite{bigann2023}. Meanwhile, we make sure that their average recall@10
ranges from 0.70 to 0.95. The specific parameters are detailed in
Figure~\ref{tab:parameter}.
For robustness evaluation, we select $\delta$ values of 0.1, 0.3, 0.5, 0.7, and 0.9.

We first present the recall distribution across indexes using CDF-style robustness curves (\S\ref{ss:eval:cdf}),
then show the robustness--recall relationship across datasets (\S\ref{ss:eval:scatter}).

\subsubsection{Recall distribution}
\label{ss:eval:cdf}

\begin{figure}
    \centering
    \includegraphics[width=\linewidth]{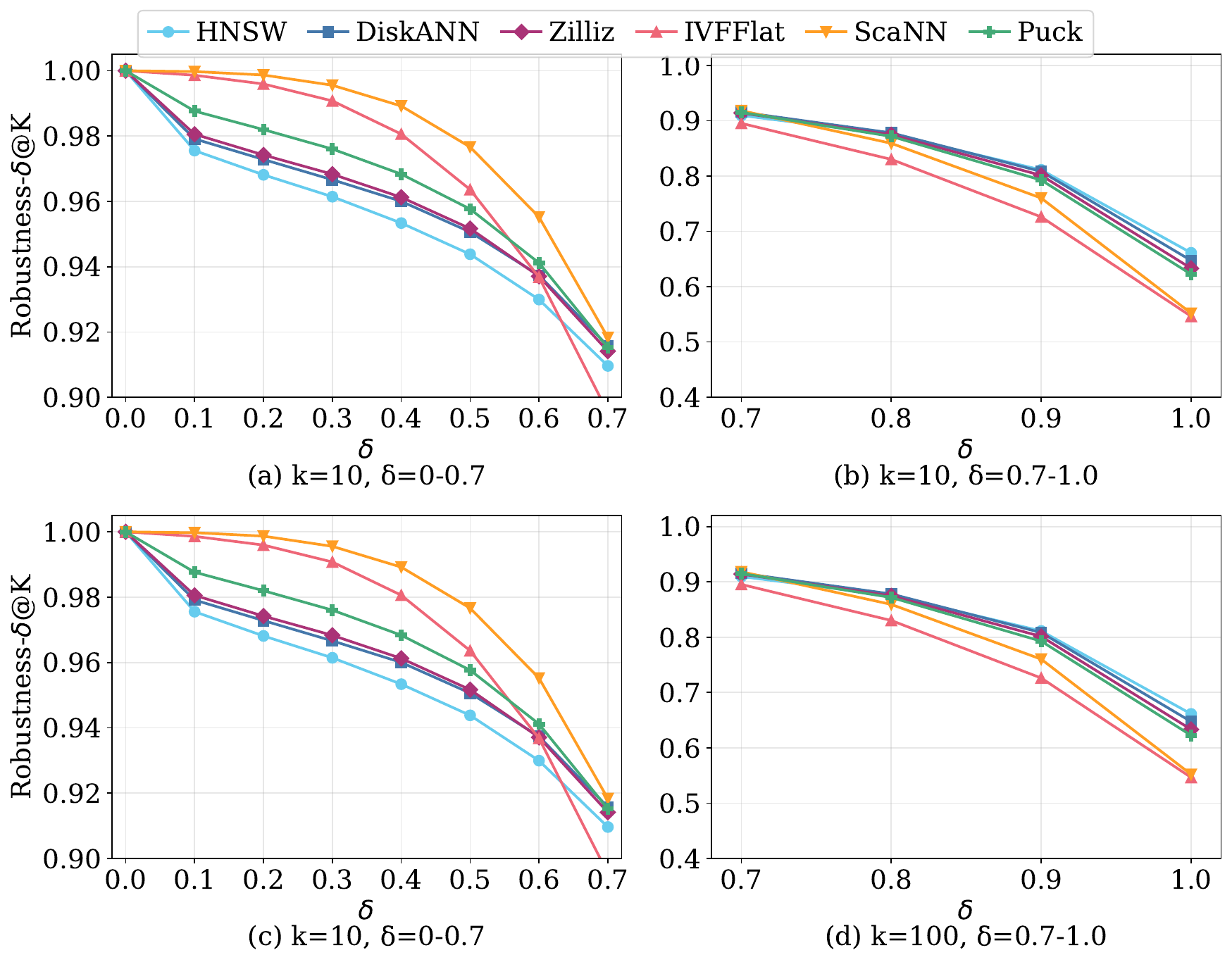}
    \vspace{-2ex}
    \caption{Index robustness on \texttoimage with $K{=}10$ and $K{=}100$.
    The x-axis represents $\delta$ values, and the y-axis shows the corresponding \robustness{} values.
    All indexes are configured to achieve the same average \recallat{10}=0.9.
    For clarity, the plot is divided at $\delta=0.7$:
    the left panel shows $\delta\le 0.7$ (note that the y-axis starts at 0.9),
    and the right one shows $\delta \ge 0.7$.
    }
    \label{fig:robustness-cdf}
    \vspace{-2ex}
\end{figure}

To study how different indexes behave at the same average recall,
we plot their robustness distributions in Figure~\ref{fig:robustness-cdf} (average \recallat{10} = 0.9).
Although their average recall is fixed, their per-query recall distributions differ substantially.
At $K{=}10$, \scann and \ivf achieve higher robustness for small $\delta$ ($\le$0.6),
indicating that most queries reach moderate recall levels.
At $K{=}100$, this advantage extends up to $\delta{\le}0.7$,
while graph-based indexes (\hnsw, \diskann) gain relatively more at higher $\delta$.

Importantly, seemingly small differences in robustness at low $\delta$ translate to large practical gaps in failure rates.
For example, on \texttoimage, \scann fails to return any relevant result for fewer than 0.03\% of queries,
whereas \diskann fails on about 2.1\%---a $70\times$ difference despite similar \robustnessat{0.1} values (0.9997 vs.\ 0.9791).
For high-recall regimes ($\delta{>}0.6$),
\hnsw and \diskann outperform partition-based ones,
achieving \robustnessat{0.9} values up to 0.81 compared to 0.73--0.76 for \ivf and \scann.

\begin{figure}
    \centering
    \includegraphics[width=\linewidth]{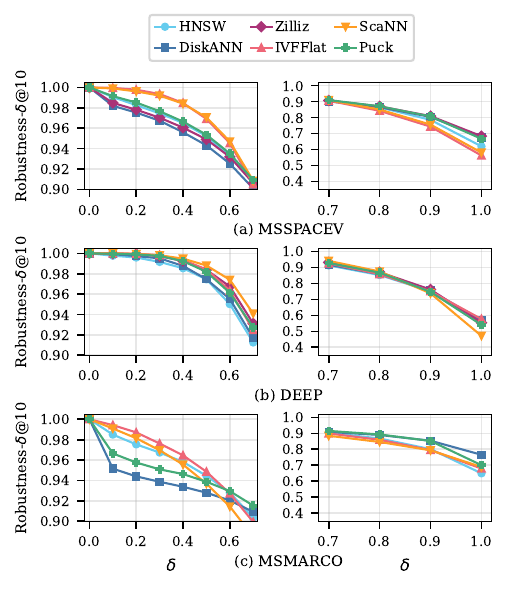}
    \vspace{-2ex}
    \caption{Index robustness on (a) \spacev, (b) \deep, and (c) \marco ($K{=}10$, average \recallat{10}=0.9).
    The same pattern holds across datasets: partition-based indexes have higher robustness at low $\delta$;
    graph-based indexes dominate at high $\delta$.}
    \label{fig:robustness-cdf-spacev}
    \label{fig:robustness-cdf-deep}
    \label{fig:robustness-cdf-msmarco}
    \vspace{-2ex}
\end{figure}

Figure~\ref{fig:robustness-cdf-spacev} shows the same analysis on \spacev, \deep, and \marco.

\paragraph{\spacev.}
The graph-vs-partition pattern holds:
\ivf and \scann achieve higher robustness at low $\delta$ ($\le$0.6),
while \hnsw and \diskann dominate at high $\delta$.
The gap between families is smaller than on \texttoimage,
because \spacev queries have less difficulty variance.

\paragraph{\deep.}
All indexes achieve high low-$\delta$ robustness ($>$0.99 at $\delta{=}0.1$),
since \deep contains fewer hard queries overall.
Differentiation appears only at $\delta{\ge}0.7$,
where graph-based indexes again show higher robustness.
\scann is an outlier with notably lower robustness at $\delta{=}1.0$,
likely due to its quantization losing precision on easy queries.

\paragraph{\marco.}
\marco (768-dimensional, inner product) reveals a richer picture.
\zilliz is excluded from \marco because its Docker image has a bug quantizing 768-dimensional vectors, producing invalid results.
\ivf and \scann maintain their strong low-$\delta$ robustness ($>$0.99 at $\delta{=}0.1$),
consistent with the partition-based pattern.
\hnsw shows a balanced distribution:
moderate at all $\delta$ levels, it avoids both extreme failures and extreme successes.
In contrast, \diskann exhibits the highest zero-recall rate (4.8\% of queries),
yet also the highest rate of perfect recalls (76.6\%), showing a strongly bimodal distribution.
Interestingly, \puck---although partition-based---behaves more like a graph index on \marco,
with a zero-recall rate of 3.4\% and high perfect-recall rate of 70.1\%.
This is because \puck uses a graph-based refinement step (tinker) on top of its partition structure;
on difficult high-dimensional queries, this graph component becomes the bottleneck,
producing the bimodal failure pattern characteristic of graph-based search.

\subsubsection{Robustness versus average recall}
\label{ss:eval:scatter}

\begin{figure*}
    \centering
    \includegraphics[width=\linewidth]{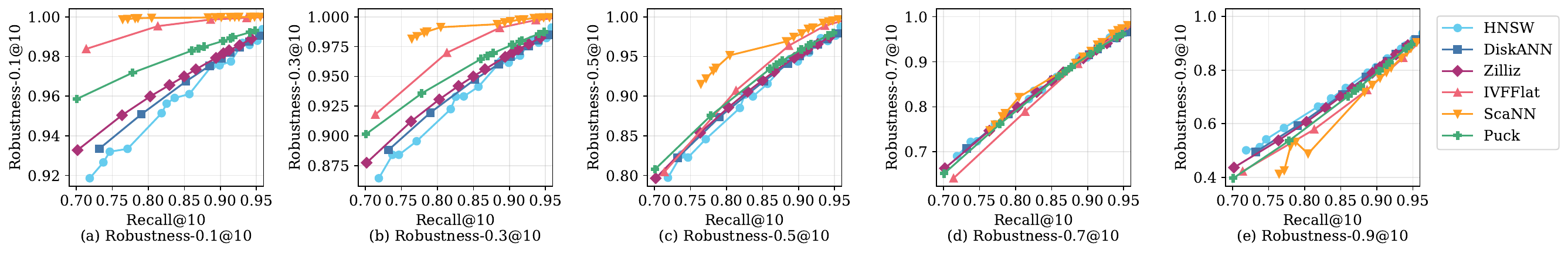}
    \vspace{-4ex}
    \caption{The \robustness{}-recall relationship of different indexes on \texttoimage. Points are operating points of the indexes. (a) shows the relationship between \robustnessat{0.3} and \recallat{10}, (b) shows the relationship between \robustnessat{0.9} and \recallat{10}.
    }
    \label{fig:text2image-overall}
    \vspace{-2ex}
\end{figure*}

To understand the relationship between robustness and recall,
we plot average recall
with \robustnessat{0.3} and \robustnessat{0.9} across indexes on the \texttoimage dataset
(Figure~\ref{fig:text2image-overall}).

At low thresholds ($\delta{=}0.3$),
partition-based indexes such as \scann and \ivf maintain consistently high robustness across recall ranges
(e.g., \robustnessat{0.3} $\approx 0.996$ at average recall 0.9),
while graph-based indexes show larger drops as average recall decreases
(e.g., \zilliz \robustnessat{0.3} $\approx 0.877$ at average \recallat{10} $\approx 0.7$,
whereas \scann still achieves \robustnessat{0.3}=0.959).
ScaNN's robustness drops more noticeably at $\delta{=}0.3$ than at $\delta{=}0.1$ as average recall decreases:
scanning fewer clusters at lower recall settings has a larger impact when the threshold
requires at least three true nearest neighbors ($\delta{=}0.3$) versus one ($\delta{=}0.1$).
In contrast, at higher thresholds ($\delta{=}0.9$), the trend reverses:
graph-based indexes achieve higher robustness (e.g., \diskann 0.81 vs.\ \scann 0.77 at average \recallat{10} of 0.9),
reflecting their strength in retrieving nearly all ground-truth neighbors when recall must be exceptionally high.

We observe similar patterns on \spacev and \deep.
On \spacev, the gap between index families is smaller because the query set
has less difficulty variance compared to \texttoimage (whose queries are out-of-distribution for the dataset).
On \deep, all indexes achieve high low-$\delta$ robustness because the dataset contains fewer hard queries;
for example, \robustnessat{0.1} for \scann reaches 0.9999 at average recall of 0.9,
while \diskann and \zilliz reach 0.999 and 0.9994.

As K increases from 10 to 100, the indexes show an increase in their robustness scores
for small $\delta$ values (e.g., 0.1 and 0.3).
The indexes have more candidates to choose from for a larger K, 
especially for those queries with low recall.
We found that graph-based indexes gain more from this increase than partition-based indexes,
since graph-based indexes can find some of the sub-optimal nearest neighbors
that are not included in the top-K results when K is small.
To the contrary, for partition-based indexes like \scann and \ivf,
larger K means more scattered candidates across the partitions,
which can lead to a higher chance of missing the ground-truth nearest neighbors.
Robustness values for larger K (e.g., K=100) are generally lower than those for smaller K (e.g., K=10) for all indexes,
indicating that as K increases, the likelihood of missing the ground-truth nearest neighbors also increases.
%

\paragraph{Root cause analysis.}
We examine several low-recall queries from \hnsw to understand this behavior.
These queries are navigated to sub-optimal nodes in the graph, and the search continues within those regions,
terminating after reaching local optima without finding the true nearest neighbors~\cite{aumuller2021role}.
In partition-based indexes,
the search process traverses centroids and visits nearby clusters,
yielding several true nearest neighbors even for difficult queries.
This structural difference explains why partition-based indexes have more stable recall distributions.

\paragraph{Summary.}
Across all four datasets, index performance for robustness
shows a consistent pattern:
partition-based indexes (\scann, \ivf) have better tail performance (low-$\delta$),
while graph-based indexes (\hnsw, \diskann) excel at high-$\delta$ robustness.
The recall distributions of \scann and \ivf are more balanced,
while graph-based indexes exhibit more skewed distributions with both more high-recall and more low-recall queries.


\subsection{A three-way trade-off: Throughput, Recall, and Robustness}
\label{evaluation:tradeoff}
\label{ss:eval:tradeoff}

\begin{figure*}
    \centering
    \includegraphics[width=\linewidth]{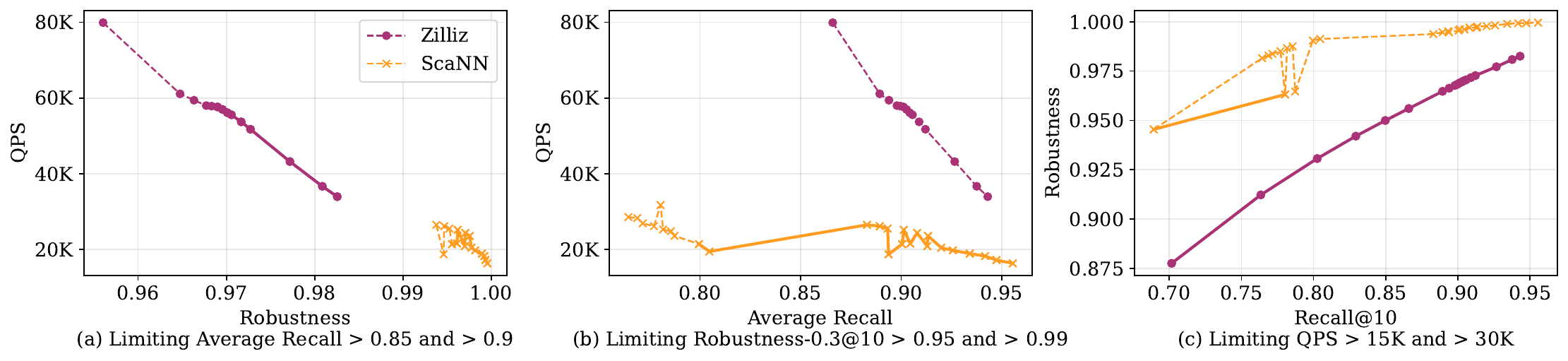}
    \caption{
    The recall-/robustness-throughput trade-off for different indexes on \texttoimage. 
    In each plot, one metric threshold is fixed, and the unsatisfied points are filtered.
    (a) Average Recall@10 is limited to no less than (1) 0.85 and (2) 0.9;
    (b) \robustnessat{0.3} is constrained to no less than (1) 0.95 and (2) 0.99;
    (c) QPS is restricted to no less than (1) 15K and (2) 30K.
    The points under threshold (1) are shown in dashed lines,
    and those with threshold (2) are shown in solid lines.}
    \label{fig:tradeoff}

\end{figure*}

Selecting an index involves a three-way trade-off among average recall, \robustness, and throughput.
The evaluation is conducted on the \texttoimage dataset.
The recall-robustness relationship is discussed in \S\ref{ss:eval:scatter}:
robustness scores tend to increase with average recall, but for small $\delta$ values,
the increases in robustness are disproportionate to the improvements in recall.

\paragraph{Guiding index selection.}
Consider a scenario where a developer needs to choose a vector index for a search application
using the \texttoimage dataset.
In a traditional setup, the only trade-off to evaluate is throughput versus recall,
where \zilliz is the clear winner: \zilliz offers better QPS than others across all recalls.
However, the selection result may change when considering robustness.
Suppose the application's users require a minimum recall of 0.3 (i.e., $\delta=$0.3).
In this case, \zilliz may not always be the optimal choice, as its \robustnessat{0.3} is sometimes outperformed by other indexes.
Next, we will elaborate on how to use \robustnessat{0.3} to select the most appropriate index.


To select the best index for a specific application,
developers should navigate the trade-off between average recall, robustness, and throughput.
A common approach is to set a threshold for one of the metrics
and then choose the candidate that offers the best trade-off point on the remaining two metrics,
provided the threshold for the first metric is satisfied.
Take the search application as an example.
If developers identify 0.99 as the threshold for \robustnessat{0.3},
they should filter out the indexes with \robustnessat{0.3}$<$0.99,
and look at the throughput-recall trade-off of the filtered results,
which is depicted in Figure~\ref{fig:tradeoff}(b).
Similar processes can be made for fixing the threshold of throughput or average recall.

Following this approach,
we conduct a comprehensive index selection for the search application example.
In Figure~\ref{fig:tradeoff}, we separately filter the results with different
thresholds of (a) average recall, (b) \robustnessat{0.3} and (c) throughput, 
and show the trade-offs of the other two metrics for the filtered results. 
We use two thresholds for each first-step filtering; we call them threshold (1) and (2).
The filtered points with threshold (1) are shown in dashed lines, and the
filtered points with threshold (2) are shown in solid lines.

In Figure \ref{fig:tradeoff}(a), the results with average recall no less than 0.85 and 0.9 are filtered, and the trade-offs of \robustnessat{0.3} and throughput are shown.
When the recall threshold is set to $>$0.85,
\zilliz can achieve the highest throughput of 80K QPS,
at the cost of lowest \robustnessat{0.3} of 0.955.
All the results (after filtering) of \scann have \robustnessat{0.3} values $>$0.996,
and the best throughput is 30K QPS.
When the recall threshold is set to $>$0.9,
\zilliz has the similar trade-off, but with fewer plausible configurations.
For a developer, they can trade off robustness for throughputs, if they care more about robustness,
they should pick \scann, otherwise \zilliz gives better throughputs.


Figure~\ref{fig:tradeoff}(b) filters by \robustnessat{0.3} $>$0.95 and $>$0.99.
At the lower threshold, \zilliz lies above \scann.
However, at $>$0.99, all \zilliz configurations are filtered out; only \scann remains.
Figure~\ref{fig:tradeoff}(c) fixes throughput at $>$15K and $>$30K.
At $>$15K, \scann always achieves higher \robustnessat{0.3} than \zilliz at the same average recall.
At $>$30K, \zilliz provides comparable \robustnessat{0.3} with higher average recall, making it a better option.
In conclusion, \scann outperforms \zilliz in some cases when making trade-offs between average recall, robustness, and throughput.
This is different from the traditional setup that ignores robustness,
in which \zilliz is always the best choice.

\subsection{End-to-End Evaluation on RAG Applications}
\label{evaluation:application}
\label{ss:eval:app}

We evaluate how vector index selection affects
Retrieval-Augmented Generation (RAG) performance,
focusing on whether \robustness provides a more
reliable assessment than average recall.
We conduct two representative RAG applications:
a \Naiverag \qa task
and an \agenticrag task.

\paragraph{Setup.}
The setup for each RAG application is summarized in Figure~\ref{tab:rag-setup}.
For each application, we evaluate multiple vector indexes,
whose configurations are detailed in Figure~\ref{tab:rag-index-config}.

\begin{figure}[t]
    \centering
    \setlength{\tabcolsep}{4pt}
\footnotesize
\begin{tabular}{@{}l|>{\centering\arraybackslash}p{3.2cm}>{\centering\arraybackslash}p{3.2cm}@{}}
  \toprule
  & \textbf{\Naiverag \qa} & \textbf{\agenticrag \qa} \\
  \midrule
  \textbf{Model} & Gemini-2.0-Flash & Search-R1 (Qwen2.5-7B), Qwen3-30B-A3B \\
  \midrule
  \textbf{Embedder} & LLM-Embedder & E5 \\
  \midrule
  \textbf{Base} & \marco (8.8M, 768d) & \wikipedia (21M, 768d) \\
  \midrule
  \textbf{Query} & \marco (6.9K) & \hotpotqa (4.8K) \\
  \midrule
  \textbf{Judger} & GPT-4o-mini & GPT-4o-mini \\
  \midrule
  \textbf{Indexes} & \knn, \hnsw, \ivf, \scann, \diskann & \knn, \hnsw, \ivf \\
  \bottomrule
\end{tabular}

    \caption{RAG application setup.
    \textbf{Models:} Gemini-2.0-Flash~\cite{gemini2}, Search-R1 (Qwen2.5-7B)~\cite{jin2025search}, Qwen3-30B-A3B~\cite{qwen2.5-7B}.
    \textbf{Embedders:} LLM-Embedder~\cite{zhang2023llmembedder}, E5~\cite{wang2022text}.
    \textbf{Datasets:} \marco~\cite{msmarco}, \wikipedia~\cite{karpukhin2020dense}, \hotpotqa~\cite{yang2018hotpotqa}.
    \textbf{Judger:} GPT-4o-mini~\cite{hurst2024gpt}. We also verify with Gemini-2.0-Flash as a secondary judge; the rankings are consistent.
    \textbf{Indexes} are the evaluated vector indexes for different RAG applications.}
    \label{tab:rag-setup}
    \vspace{-2ex}
\end{figure}

\begin{figure}[t]
    \centering
    \resizebox{\columnwidth}{!}{%
    \scriptsize
    \begin{tabular}{@{}l|cc@{}}
        \toprule
        & \textbf{\Naiverag} & \textbf{\agenticrag} \\
        \midrule
        $K$ & 10 & 5 \\
        \recallatk range & 0.85--0.95 & 0.8--0.9 \\
        \midrule
        \hnsw & M=16, efC=300 & M=64, efC=300 \\
              & efS=30--155 & efS=30--80 \\
        \midrule
        \ivf & nl=4K, np=20--100 & nl=10K, np=60--200 \\
        \midrule
        \scann & \#l=10K, ro=150, ls=10--90 & -- \\
        \midrule
        \diskann & R=48, L=500, Ls=15--60 & -- \\
        \bottomrule
    \end{tabular}}
    \caption{Index configurations for RAG applications.
    efC/efS: construction/search list size;
    nl/np: clusters/probed clusters;
    \#l/ro/ls: leaves/reorder/leaves to search;
    R/L/Ls: degree/candidate list/beam width.}
    \label{tab:rag-index-config}
    \vspace{-2ex}
\end{figure}

\begin{figure}[t]
    \centering
    \includegraphics[width=\linewidth]{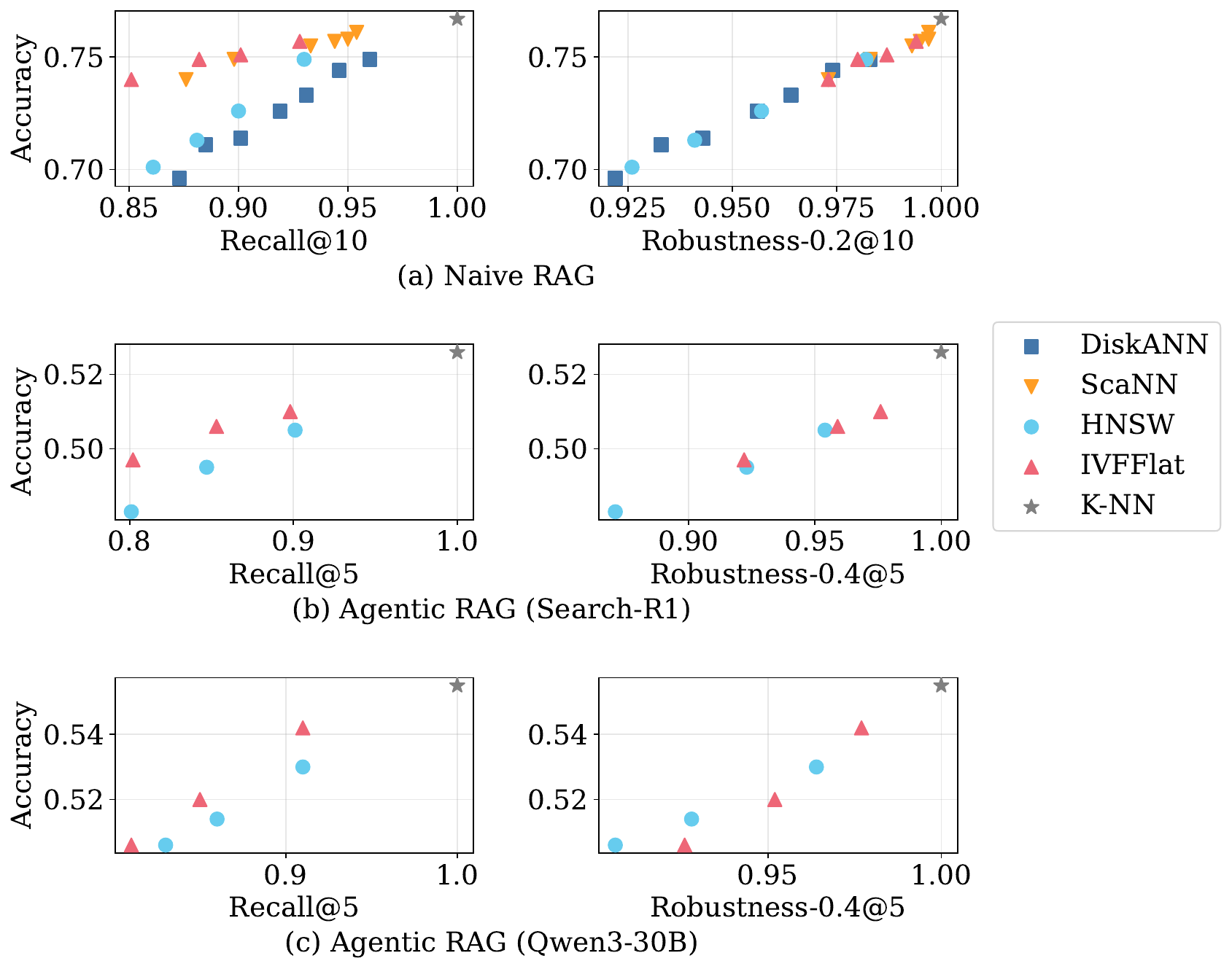}
    \vspace{-4ex}
    \caption{End-to-end RAG accuracy versus robustness and recall.
    \textbf{Left}: accuracy as a function of recall.
    \textbf{Right}: accuracy as a function of robustness.
    Different colors represent different indexes.
    }
    \label{fig:qa-accuracy}
    \vspace{-2ex}
\end{figure}

\paragraph{\Naiverag \qa.}
Each question in the \marco dataset can be directly answered by a single relevant document,
making the task primarily dependent on retrieval quality.
We use Gemini-2.0-Flash~\cite{gemini2} as the LLM, with LLM-Embedder~\cite{zhang2023llmembedder} for encoding.
The LLM answers 76.7\% of questions correctly on brute-force \knn search results,
representing the upper bound of achievable performance.
We select $\delta=0.2$ based on sampling: questions with \recallat{10}=0.4 maintain accuracy over 91\% of \knn;
at \recallat{10}=0.2, it drops to 70\%; at 0.1, only 40\%.
Figure~\ref{fig:qa-accuracy}(a) shows that partition-based indexes (\scann, \ivf) consistently achieve higher \qa accuracy than
graph-based ones (\diskann, \hnsw) at similar average recall levels.
A \scann configuration with \recallat{10} = 0.90 achieves the same \qa accuracy
as a \diskann configuration with \recallat{10} = 0.96 (74.9\%), while providing significantly higher throughput (25,054 vs. 14,606 QPS).
The right subfigure shows that \qa accuracy correlates more strongly
with \robustnessat{0.2} than with average recall across all indexes.

\paragraph{\agenticrag.}
We adopt the Search-R1~\cite{jin2025search} framework, which uses reinforcement learning
to train an LLM to decompose complex questions and issue retrieval tool calls.
We evaluate two models: the original Search-R1 (Qwen2.5-7B)~\cite{qwen2.5-7B}
and a larger Qwen3-30B-A3B, both on \hotpotqa~\cite{yang2018hotpotqa}.
We integrate \hnsw and \ivf as retrieval backends with $K=5$.
Using sampling, we find that queries with \recallat{5} $<$ 0.4 mark a phase change in accuracy.
Figure~\ref{fig:qa-accuracy}(b) shows that with Search-R1 (Qwen2.5-7B),
\ivf at average \recallat{5} = 0.8 achieves similar end-to-end accuracy to \hnsw at a much higher average \recallat{5} = 0.9.
Figure~\ref{fig:qa-accuracy}(c) confirms the same pattern with the larger Qwen3-30B-A3B model,
which achieves higher absolute accuracy but exhibits the same index-dependent behavior:
partition-based indexes (\ivf) outperform graph-based indexes (\hnsw) at comparable average recall.

However, the larger Qwen3-30B-A3B model partially masks retrieval failures
by falling back to internal knowledge.
Using an LLM judge, we find that even with \knn (perfect retrieval),
233 correct answers rely on internal knowledge---a baseline
where the corpus genuinely lacks the needed information.
\hnsw at \recallat{5}=0.91 adds 25 additional internal-knowledge answers
(258 total) due to its retrieval failures,
while \ivf at the same recall adds nearly none,
reflecting its higher \robustnessatk{0.4}{5}.
When we enforce retrieval-only answers (excluding internal knowledge),
the accuracy becomes 55.5\% (\knn), 54.2\% (\ivf),
and 53.0\% (\hnsw).
We also observe that over 60\% of questions trigger retry queries
when the initial retrieval fails,
but two-thirds of these retries still fail to retrieve relevant documents,
indicating that index-level failures are difficult to recover from.
The smaller Search-R1 model shows a larger gap between indexes
because it cannot fall back to internal knowledge,
suggesting that \robustness is especially critical
for retrieval-dependent models and private-domain applications.

\paragraph{Summary.}
Across both RAG applications, partition-based indexes consistently achieve better accuracy
than graph-based indexes at the same average recall, due to their superior low-$\delta$ robustness.
\robustness serves as a better predictor of end-to-end RAG performance than average recall alone.

\subsection{\robustness for Different Index Families}
\label{ss:eval:index}

Prior experiments (\S\ref{evaluation:overall}),
show a substantial difference in the robustness characteristics
between graph-based indexes~\cite{wang2021graphann} (\hnsw, \diskann and \zilliz) and partition-based indexes (\scann, \ivf and \puck).
Note that \puck is a hybrid: it uses partitioning with a graph-based refinement step,
so it can exhibit characteristics of both families depending on the dataset and query difficulty.
The IVF-based indexes usually have higher robustness values than the graph-based ones when the $\delta$ values are small ($<$0.5),
whereas the graph-based indexes have higher robustness values when the $\delta$ value is large ($>$0.8).

In this section, we conduct an in-depth study on the recall distribution of the graph-based indexes and the IVF-based indexes,
and analyze the impact of the index structure, parameters, and techniques on the recall distribution of these two families of vector indexes.
The evaluation is conducted based on two baseline indexes: \hnsw for
the graph-based index family and \ivf for the IVF-based index family, both implemented in the Faiss library. 
We use the \texttoimage dataset.

\begin{figure}
    \centering
    \includegraphics[width=\linewidth]{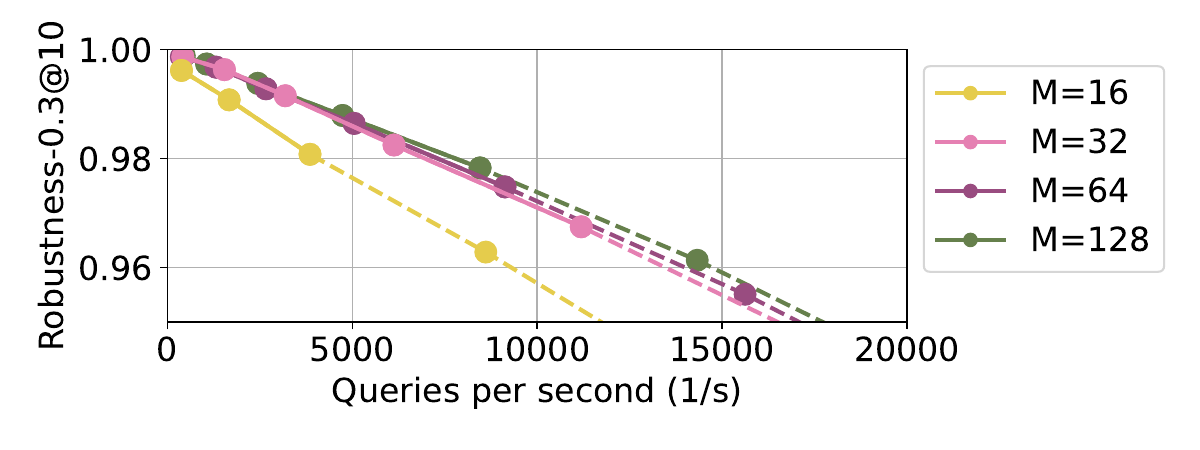}
    \caption{The \robustnessat{0.3} of \hnsw with different parameter settings on \texttoimage.
    Each line represents the \robustnessat{0.3} of \hnsw with different \textbf{M} values.
    Each point on a line is a configuration with different \textbf{efSearch} values, \textbf{efSearch} changes from 128 to 4096.
    We use an average recall of 0.9 as the threshold.
    Points meeting this threshold are depicted in solid lines,
    while those that do not are displayed in dashed lines.
    }
    \label{fig:parameter-study-hnsw}
    \vspace{-10pt}
\end{figure}

\subsubsection{Graph-based Index}
\label{evaluation:graph}

We tune the parameters of \hnsw and observe the change of average recall@10, \robustnessat{0.3} and \robustnessat{0.9} as the throughput changes.
Three critical tuning parameters of \hnsw are \emph{M}, \emph{efSearch} and \emph{efConstruction}.
\emph{M} is the maximum number of neighbors to keep for each node in the graph, and \emph{efSearch} is the number of neighbors to visit during the search process.
\emph{efConstruction} is the number of candidate neighbors to visit during the graph construction process.
As long as \emph{efConstruction} is set to a value larger than \emph{M}, it will not affect the search performance.

As shown in Figure~\ref{fig:parameter-study-hnsw},
we plot the \robustnessat{0.3}-throughput curves of \hnsw with different \emph{M} values and \emph{efSearch} values.
Each line represents the \robustnessat{0.3} of \hnsw with different \emph{M} values, and each point on a line is a configuration with different \emph{efSearch} values.
We found that the \robustnessat{0.3} of \hnsw is highly related to the \emph{efSearch} value
when the \emph{M} is above a certain value (32 in this case).

The \robustnessat{0.3} of \hnsw can reach 0.9988 when the \emph{efSearch} is set to 4096 and \emph{M} is set to 128.
Under this setting, there is a big set of candidate neighbors to visit during the search process,
which can help the search process escape the local optima and find the true nearest neighbors of a query.
However, the cost of achieving such high \robustnessat{0.3} is the throughput dropping to only 451 QPS, 
an unacceptable performance for most of the use cases.
As for the setting of \emph{M}, \robustnessat{0.3} lines with different \emph{M} values are close to each other when the \emph{M} is set to 32, 64, and 128.
It is hard for \hnsw to achieve high \robustnessat{0.3} when the throughput is high even for a very large \emph{M} value.
While if the \emph{M} is set to 16, the \robustnessat{0.3} of \hnsw is lower than that of other configurations under a similar throughput.
For example, when the throughput of \hnsw reaches around 4K QPS, the
\robustnessat{0.3} of \hnsw with \emph{M} = 16 is 0.98, and the
\robustnessat{0.3} of \hnsw with \emph{M} = 128 is roughly 0.99.
This is because the number of neighbors to keep for each node in the graph is small,
and the search process may be navigated to a sub-optimal node in the graph.

It is important that the index user should choose the \emph{M} value based on the requirements of their applications.
Since a higher \emph{M} requires more memory and a longer index construction time,
and the benefit of a larger \emph{M} is not significant when it is above a certain value (e.g., 32).

\paragraph{A closer look at the low-recall queries.}
To study the cause of low recall,
we select a few low-recall queries from \hnsw with \text{efSearch}=4096.
These queries are navigated to sub-optimal nodes in the graph, and perform further searches in those sub-optimal areas,
and stop after they hit the local optima.
So they fail to find the true nearest neighbors for the query vectors.

To quantify the severity of these failures,
we examine which ground-truth neighbors the retrieved vectors correspond to.
For each retrieved vector, we check its rank in the ground-truth top-100 list
(rank 1 = closest neighbor).
On \texttoimage, for queries where \hnsw achieves recall $\le 0.3$,
over 41\% of retrieved vectors fall \emph{outside} the ground-truth top-100 entirely,
and only 9\% are among the 10 closest neighbors.
In contrast, \ivf under the same failure condition retrieves vectors that are
still relatively close: 22\% are among the 10 closest neighbors,
and less than 2\% fall outside the top-100.
The pattern is consistent across all four datasets.
This means graph-based failures are qualitatively worse:
the index does not merely miss some neighbors,
it returns vectors from a completely different region of the space.
For applications like RAG, where retrieved document relevance directly
affects output quality, this distinction matters.
A common mitigation is to retrieve more candidates (e.g., $K{=}100$) and rerank,
but at the efSearch values typically used in practice ($1.5$--$5\times K$),
over half of severe failures still do not recover the closest ground-truth neighbor.



\begin{figure}
    \centering
    \includegraphics[width=\linewidth]{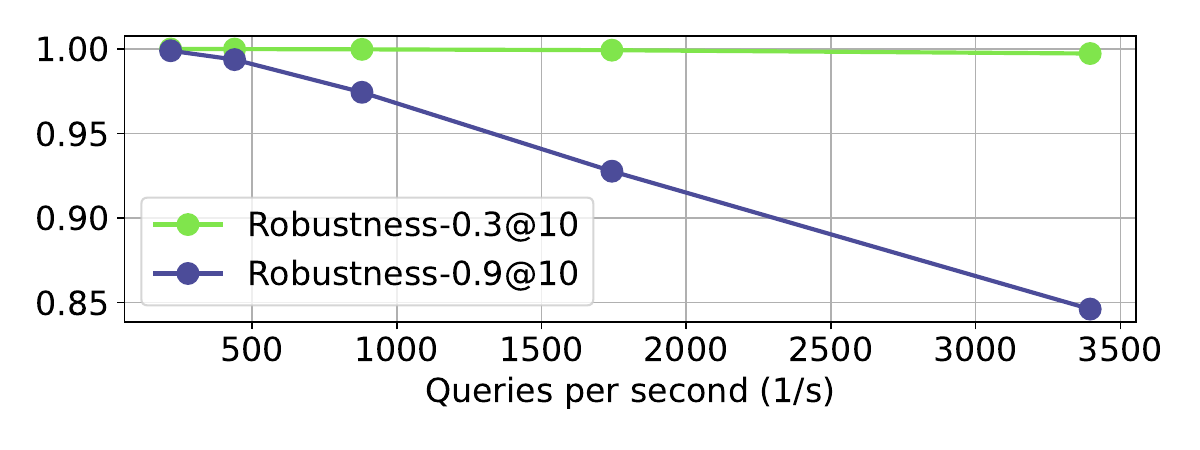}
    \caption{The \robustnessat{0.3} and \robustnessat{0.9} of \ivf with different parameter settings on \texttoimage.
    Each point on the line is a configuration with different \textbf{n\_probe} values, \textbf{n\_probe} changes from 64 to 1024.
    \textbf{n\_list} is fixed to 10K.
    Points with average recall$<$0.9 are filtered.}
    \label{fig:parameter-study-ivfflat}
    \vspace{-10pt}
\end{figure}

\subsubsection{Partition-based Index}
\label{evaluation:ivf}

We tune the parameters of \ivf and observe the change of average recall@10, \robustnessat{0.3} as the throughput changes.
The critical tuning parameter of \ivf is \emph{n\_probe}, which is the number of clusters to visit during the search process.
As shown in Figure \ref{fig:parameter-study-ivfflat}, we plot the \robustnessat{0.3}-throughput
and \robustnessat{0.9}-throughput curves of \ivf with \emph{n\_list}=10K, meaning partitioning the dataset into 10K clusters. 
Each point on the line is a configuration with different \emph{n\_probe} values.
We find that the \robustnessat{0.3} of \ivf is stable.
When the throughput of \ivf changes significantly (218 $\to$ 3,395 QPS),
the \robustnessat{0.3} drops only slightly (0.99997 $\to$ 0.9975).
In contrast, the \robustnessat{0.9} of \ivf shows a substantial drop (0.9985 $\to$ 0.84 ).

We further analyze the \ivf with the \emph{n\_probe} to be small numbers. 
We observe that by varying \emph{n\_probe} from 1 to 8,
\robustnessat{0.3} of \ivf increases significantly from 0.46 to 0.91.
Similarly, the corresponding \robustnessat{0.1} raises from 0.75 to 0.98.
These numbers show that the IVF-based index is able to find at least one true nearest neighbor for most queries
by just scanning less than 1/1000 of the clusters in the search process.
This is important for applications that require high robustness and high throughput.

\section{Lessons Learned}
\label{s:lessons}
\label{s:lesson}


Below, we summarize the key observations from our evaluation
and offer guidance on applying \robustness.
We aim to provide practical insight for its future use.

\heading{Key observations.}
\label{s:lessons-observation}
Based on our evaluation, we summarize five key observations:

\vspace{0.5ex}
\noindent
\emph{1. Indexes exhibit significant differences in recall distributions,
  despite having the same average recall.}
This is confirmed in Figure~\ref{fig:recall-dist},
\ref{fig:robustness-cdf}, \ref{fig:text2image-overall},
and several other experiments (\S\ref{evaluation:overall}).
This observation motivates the need for a new metric to
comprehensively evaluate vector indexes.

\vspace{0.5ex}
\noindent
\emph{2. \robustness enables a more comprehensive evaluation of
vector indexes than existing metrics.}
As illustrated in section~\ref{s:metric}, standard metrics fail to fully
characterize vector index behavior.
In contrast, \robustness, parameterized by $\delta$, captures both low- and high-recall tail behaviors.

\vspace{0.5ex}
\noindent
\emph{3. \robustness aligns with end-to-end application targets,
with an appropriate $\delta$.}
As shown in \S\ref{ss:eval:app}, \robustnessatk{0.2}{10} 
serves as a
good predictor for RAG Q\&A accuracy, demonstrating
its effectiveness in capturing application-level performance requirements.

\vspace{0.5ex}
\noindent
\emph{4. Two mainstream index families exhibit structural
differences in recall behavior.}
Graph-based indexes---including \hnsw, \diskann, and \zilliz---tend to produce
skewed recall distributions, with more queries at both the high
and low ends. Therefore, they often perform better on high-$\delta$
\robustness but worse on low-$\delta$ ones. In contrast, partition-based
indexes such as \ivf and \scann typically exhibit more uniform recall
distributions across queries.

\vspace{0.5ex}
\noindent
\emph{5. Tuning index parameters can improve \robustness.}
For graph-based indexes, parameters such as the graph maximum degree (e.g., \emph{M} in \hnsw)
influence their \robustness. Higher degrees lead to better connectivity, thus better worst-case performance.
For partition-based indexes, parameters like the number of clusters searched
(e.g., \emph{n\_probe} in \ivf) affect the \robustness, particularly for high-$\delta$ ones.

\label{s:lessons-guidelines}
\heading{Choosing vector indexes using \robustness.}
Based on our observations, we offer practical guidelines for using \robustness
to select vector indexes.

\vspace{0.5ex}
\noindent
\emph{1. Choosing an index family based on \robustness.}
By observation 4,
for applications requiring high-$\delta$ robustness,
graph-based indexes are generally more suitable.
In contrast, partition-based indexes work better for applications prioritizing
low-$\delta$ robustness.

\vspace{0.5ex}
\noindent
\emph{2. Selecting $K$ and $\delta$.}
There is no one-size-fits-all solution for choosing $K$ and $\delta$.
The value of $K$ is typically determined by application
and reflects how many results are needed for downstream tasks.
Larger $K$ allows room for post-processing (e.g., reranking, filtering) but incurs high query cost.
The choice of $\delta$ depends on application requirements.
Applications generally fall into two categories:
(i) low-$\delta$ preference: applications tolerate some irrelevant results as
long as enough relevant items are retrieved (e.g., RAG Q\&A)
and (ii) high-$\delta$ preference: applications require strict correctness,
where even a few incorrect items are unacceptable (e.g., exact-match
recommendations~\cite{hou2024bridginglanguageitemsretrieval}).


\vspace{0.5ex}
\noindent
\emph{3. Balancing Average Recall, \robustness, and Throughput.}
As illustrated in Section~\ref{ss:eval:tradeoff},
selecting an index and its configuration involves a three-way trade-off.
Developers should fix one metric and explore trade-offs between the other two.
For example, by fixing throughput, one can plot the trade-off curve between
average recall and \robustness, then choose an index configuration that balances overall
accuracy and tail robustness for the application.

\heading{Improving \robustness.}
Beyond choosing different indexes and tuning parameters,
\robustness can be improved by applying additional techniques or reconstructing indexes.
We list several approaches below.

\vspace{0.5ex}
\noindent
\emph{Applying product quantization (PQ).}
PQ compresses vector representations and is originally
designed to reduce memory and computation costs.
We observe that when applying PQ, expanding the number of candidates (e.g., the
reorder parameter in \scann) improves both average recall and \robustness at high-$\delta$.

\vspace{0.5ex}
\noindent
\emph{Adaptive parameter tuning during the search process.}
Learned prediction models can adaptively select optimal search parameters for
each query based on query features and runtime signals. Originally proposed to
improve search efficiency~\cite{li2020improving}, this approach also improves
robustness by stabilizing the recall distribution: it allocates more resources
to difficult queries with low recall and fewer to easier ones, leading to more
consistent performance across queries.

\vspace{0.5ex}
\noindent
\emph{Training index with the query set.}
Prior research has explored training indexes with query sets
such as adding extra edges in a graph-based index~\cite{chen2024roargraph}
and replicating vectors into multiple clusters for a partition-based index~\cite{sun2024soar}.
These approaches improve \robustness at high-$\delta$ thresholds by increasing the likelihood
that outliers or difficult queries retrieve accurate results.

\section{Conclusion}

This paper introduces \robustness, a new metric that captures recall distribution
against application-specific threshold $\delta$, addressing the limitations of average recall.
It offers a clearer view of retrieval quality, especially
for tail queries that impact end-to-end performance.
By integrating \robustness into standard benchmarks,
we reveal substantial robustness differences across today's vector indexes.
We also identify design factors that influence robustness and provide
practical guidance for improving robustness for real-world applications.

\bibliographystyle{plain}
\bibliography{ref}{}

\end{document}